\documentclass[prb,aps,amsmath,showpacs,preprint]{revtex4}
\usepackage{graphicx}

\usepackage{dcolumn}

\usepackage{bm}

\begin{document}
\title{Ground-state valency and spin configuration of the Ni-ions in nickelates}
\author{L. Petit$^{1}$, G. M. Stocks$^{2}$, T. Egami$^{2,3}$, Z. Szotek$^{4}$, and W.M. Temmerman$^{4}$}
\affiliation{
$^{1}$ {\it Computer Science and Mathematics Division, and Center for Nanophase Materials Sciences,}
{\it Oak Ridge National Laboratory, Oak Ridge, TN 37831, USA} \\
$^{2}$ {\it Materials Science and Technology Division, Oak Ridge National Laboratory,}
{\it Oak Ridge, TN 37831, USA} \\
$^{3}$ {\it Department of Materials Science and Engineering}
{\it and Department of Physics and Astronomy,}
{\it University of Tennessee, Knoxville, Tennessee 37996, USA }\\
$^{4}$ {\it Daresbury Laboratory, Daresbury, Warrington WA4 4AD, UK} }

\date{\today}

\begin{abstract}

The ab initio self-interaction-corrected local-spin-density approximation is used
to study the electronic structure of both stoichiometric and non-stoichiometric nickelates.
From total energy considerations it emerges that, in their ground-state, both
LiNiO$_2$ and NaNiO$_2$ are insulators, with the Ni ion in the Ni$^{3+}$ low spin state
($t_{2g}^6e_{g}^1$) configuration.
It is established that a substitution of a number of Li/Na atoms by divalent impurities
drives an equivalent number of Ni ions in the NiO$_2$ layers from the JT-active trivalent low-spin
state to the JT-inactive divalent state. 
We describe how the observed considerable differences between LiNiO$_2$ and NaNiO$_2$
can be explained through the creation of Ni$^{2+}$ impurities in LiNiO$_2$.
The indications are that the random distribution of the Ni$^{2+}$ impurities might be 
responsible for the destruction of the long-range orbital ordering in LiNiO$_2$.
\end{abstract}


\maketitle

The metal oxide compounds LiMO$_2$ (M=Ni, Co, Mn) are of considerable interest due to 
their potential application as a cathode material in lithium-ion rechargeable batteries~\cite{mizushima}. 
In particular, LiNiO$_2$ and its isoelectronic counterpart NaNiO$_2$ have been the subject of numerous
model and first-principles calculations focussing on the possible charge, spin and orbital degrees of
freedom, and their impact on the electronic and magnetic properties of these compounds.
In most model calculations, a JT-active Ni$^{3+}$ ion with $t_{2g}^6$e$_g^1$ ground-state 
configuration has been assumed, in agreement with experimental evidence 
~\cite{chappelEur615}, although a Ni$^{2+}$($d^{8}$)
ion has also been suggested~\cite{kuiper}.
With the Ni$^{3+}$ configuration, model calculations have been able to reproduce the experimentally observed orbital
ordering in NaNiO$_2$ and for LiNiO$_2$, a number of models have been proposed to explain the
absence of long range magnetic order (see for example Reynaud {\it et al.}~\cite{reynaud} and references therein).
However, no universally accepted model has emerged that can simultaneously explain magnetic and orbital ordering in NaNiO$_2$ and the absence of the very same in LiNiO$_2$.
First-principles band structure calculations based on the local spin density (LSD) approximation 
to density functional theory (DFT)
have not even been able to yield the expected insulating ground state~\cite{aydinol}.
Assuming a low-spin ground-state configuration for the Ni$^{3+}$-ion, electronic structure calculations, 
using the LDA+U approach, 
result in the opening of an insulating gap in both LiNiO$_2$~\cite{anisimov} and NaNiO$_2$~\cite{meskine}.
Although very different in nature from model approaches, the LDA+U method also relies on the introduction of the Hubbard U to
account for the strong correlations. As there is no straightforward approach for determining the U from first-principles,
it is usually treated as a parameter in the calculations.

The motivation for the present letter is to apply the self-interaction corrected (SIC) LSD method and
seek first-principles, parameter free, understanding of the ground state spin and valence configurations of nickelates.
Based purely on the total energy considerations, we show that Ni$^{3+}$ $t_{2g}^6$e$_g^1$ is the ground-state 
configuration in the stoichiometric compounds.
Furthermore we predict valency changes upon substitution of Li for Ni with potentially far reaching
consequences for the magnetic and orbital odering.

The SIC-LSD method~\cite{svane,temmerman} is an {\it ab initio} approach that corrects for an
unphysical self-interaction contained in the LSD total energy functional~\cite{perdew}. 
While, for extended band states, the resultant error in the LSD energy is generally insignificant, 
it may be considerable for atomic-like localized states. In the SIC-LSD method both localized and 
delocalized states are expanded in the same set of basis functions, and are thus treated on an equal footing.
Different localized/delocalized configurations are realized by assuming different numbers of localized
states - here $d$-states on Ni-atom sites. Since the different localization scenarios constitute distinct
local minima of the same energy functional, $E^{SIC}$, their total energies may be compared and 
the global energy minimum then defines the ground state total energy {\em and} the valence configuration
of the Ni-ion. This latter is defined as the integer number of electrons available for band formation, namely
\begin{equation}
N_{val}=Z-N_{core}-N_{SIC}, 
\end{equation}
where Z is the atomic number (28 for Ni), $N_{core}$ is the number of core (and semicore) electrons (18 for Ni), and $N_{SIC}$ is the number of localized, i.e. self-interaction corrected, states.


LiNiO$_2$ crystallizes in a trigonal structure ($R\bar{3}m$ (166) space group) that consists of layers of NiO$_2$ slabs built from edge sharing NiO$_6$ octahedra and separated from each other by a layer of Li cations~\cite{rougier,hewston} (see Fig. 1).
\begin{figure}[h!]
\includegraphics[scale=0.40,angle=0,clip]{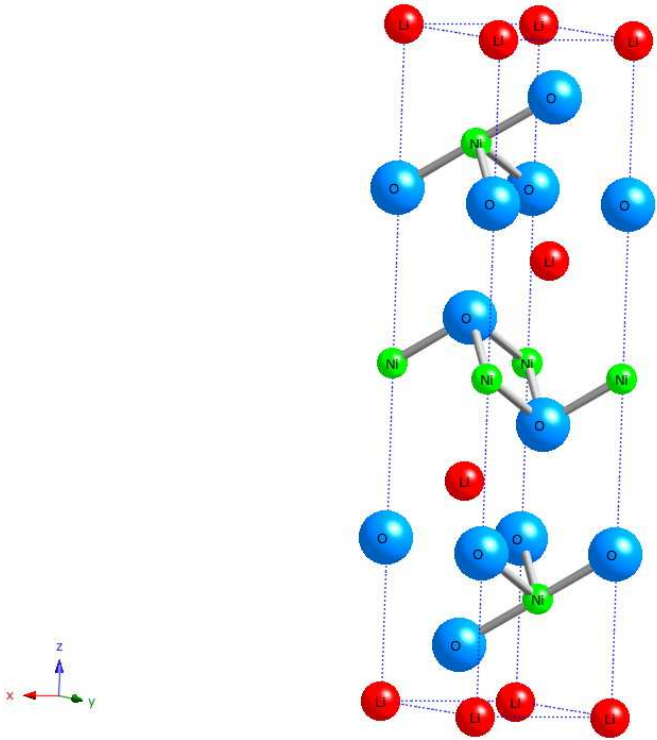}
\caption{
LiNiO$_2$ crystal structure, consisting of alternating layers of O (blue), Li (red), and Ni (green) atoms.
\label{crystal}
}
\end{figure}
This crystal structure, which can be derived from the rock salt structure of NiO by replacing every second Ni (111) plane by a Li plane, is favorable
to fast-ion intercalation/removal of Li ions during a battery discharge/charge cycle. NaNiO$_2$ is isoelectronic to LiNiO$_2$ and, at high
temperatures, also crystallizes in the $R\bar{3}m$ structure. However, below T=480 K, NaNiO$_2$ undergoes a crystal structure distortion from
rhombohedral to monoclinic (C2/$m$ (12) spacegroup) that is driven by a cooperative Jahn-Teller
ordering in the Ni-ion layer~\cite{chappelEur615}. In LiNiO$_2$, a local distortion of the NiO$_6$ tetrahedra has been observed~\cite{rougier} and
associated with the expected Jahn-Teller (JT) activity of the Ni ions in an octahedral environment. However, unlike NaNiO$_2$, the crystal
structure shows no indication of collective JT-distortion. It has been suggested,
that this is due to the formation of three sublattices with non-collinear JT-axes, and the strain fields
generated by the triangular symmetry of the NiO$_2$ slabs prevent the local orbitral ordering from developing into long-range ordering ~\cite{chung}.
The difference in magnetic behaviour is equally puzzling in that the observed magnetic order of NaNiO$_2$,
ferromagnetic intra-(Ni)layer coupling with anti-ferromagnetic inter-(Ni)layer coupling~\cite{bongers,lewis},
does not seem to occur in LiNiO$_2$.

\begin{figure}[h!]
\includegraphics[scale=0.60,angle=-90,clip]{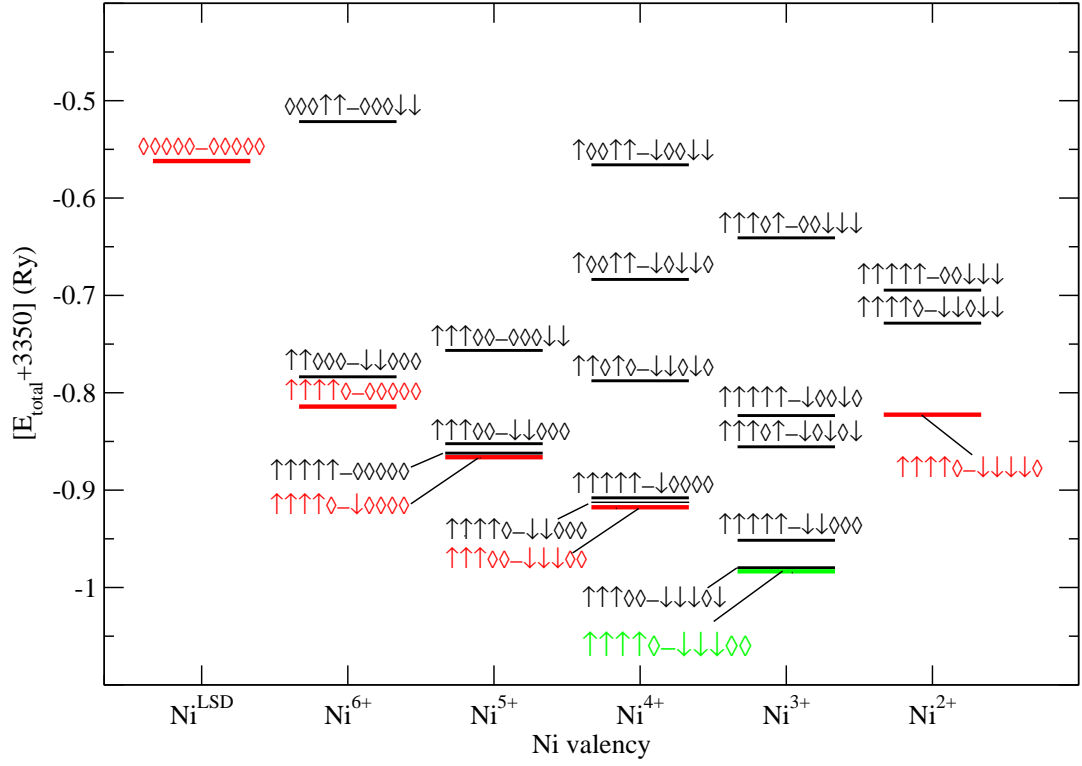}
\caption{
Total energy as a function of Ni $d$-configuration in LiNiO$_2$. A number of scenarios are shown for
each given configuration, $d_{xy}^{\uparrow}d_{yz}^{\uparrow}d_{xz}^{\uparrow}d_{3z^2-r^2}^{\uparrow}
d_{x^2-y^2}^{\uparrow}-
d_{xy}^{\downarrow}d_{yz}^{\downarrow}d_{xz}^{\downarrow}d_{3z^2-r^2}^{\downarrow}d_{x^2-y^2}^{\downarrow}$,
where a localized spin up/spin down state is indicated by the ${\uparrow}/{\downarrow}$ symbol, and a delocalized
state is indicated by the $\diamondsuit$ symbol.
\label{Eplot}
}
\end{figure}
Assuming the undistorted $R\bar{3}m$ crystal structure for LiNiO$_2$, we have calculated the total energies for five Ni valences -- Ni$^{2+}$ through Ni$^{6+}$ -- as well as the completely delocalized LSD configuration.  To determine the lowest total energy for each Ni valency (i.e. number of localized d-states), we have considered several possible ways the localized states can be permuted among the available $d$-orbitals. The results are summarized in Fig. \ref{Eplot}. Clearly, the LSD is the least favorable scenario. Overall, the different valency configurations become energetically more favorable as the number of localized $d$-electrons increases. The global energy minimum is obtained for the trivalent Ni$^{3+}$ scenario with seven localized $d$-electrons in the t$_{2g}^6$e$_g^1$ low spin configuration. This is marked by the green line in Fig. \ref{Eplot}. Compared to Ni$^{3+}$, the Ni$^{2+}$ configuration is again less favorable. Based on total energy considerations,
the SIC-LSD thus validates the choice of charge and spin configuration used in the model and LDA+U calculations of LiNiO$_2$. The same groundstate configuration is found in a similar study of the electronic structure of NaNiO$_2$.

\begin{figure}[t!]
\begin{center}
\includegraphics[scale=0.50,angle=-90,clip]{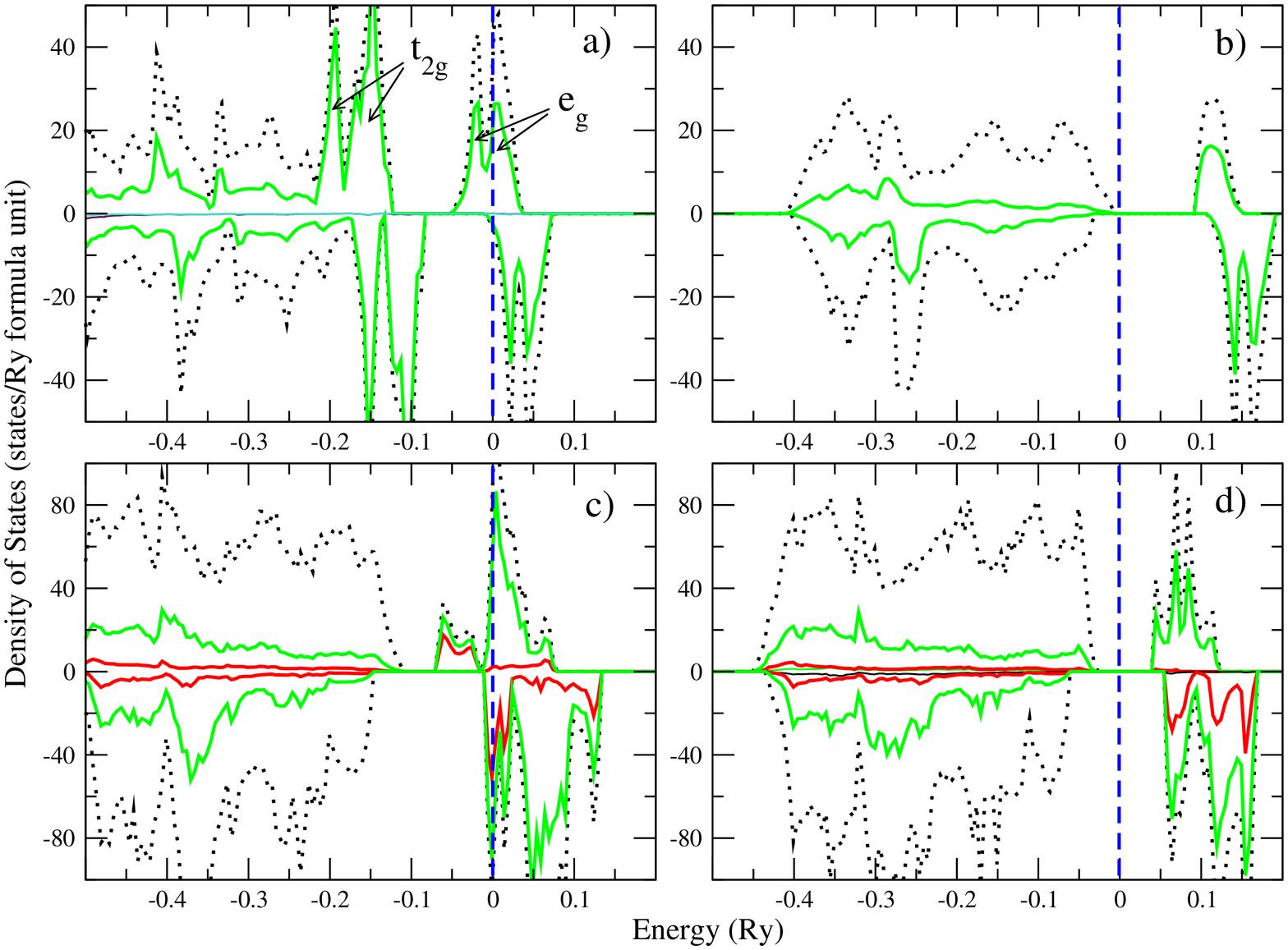}
\end{center}
\caption{
Spin-decomposed density of states as obtained for a) LiNiO$_2$ with the LSD, b) LiNiO$_2$ with the Ni$^{3+}$ groundstate configuration, c)
Li$_3$Ni$_{Li}^{3+}$Ni$_3^{3+}$O$_8$, and d) Li$_3$Ni$_3^{3+}$Ni$_{Li}^{2+}$Ni$^{2+}$O$_8$.
The green and black lines represent respectively the Ni $d$-projected and total DOS. The red in c) and d)
represents the $d$-projected DOS of the Ni substituted in the Li layer. The perpendicular dashed blue line indicates the position of the Fermi level.
\label{lsddos}
}
\end{figure}
In actuality an indication of the SIC-LSD preference for the low spin ground state of the Ni-ion is already inherent 
in the LSD calculation. As can be seen from
the  density of states (DOS) shown in Fig. \ref{lsddos}a, the splitting of $\sim 2.0$ eV between the majority Ni $t_{2g}$ and
$e_g$ states (crystal field splitting)
is some four times larger than the exchange splitting ($\sim 0.5$ eV) between majority and
minority Ni $d$-states.
A detailed inspection of this same DOS reveals that the $t_{2g}$ states
in both spin channels are fully
occupied while the majority $e_g$ states are only partially filled, as the Fermi energy falls in 
the range of hybridized O 2$p$ and Ni $e_g$ states.
Since the self-interaction is only sizeable for the occupied and fairly localized states, one can easily see
that only these occupied Ni $d$-states will benefit from the self-interaction correction,
thus leading to the global energy minimum of the SIC-LSD energy functional for this low-spin state.

When comparing the SIC-LSD DOS for the ground state configuration (Fig. \ref{lsddos}b) to the LSD DOS
(Fig. \ref{lsddos}a), one notices that the $t_{2g}$ states at the top of the valence band and parts 
of the spin-up $e_g$ band states at the bottom of the
conduction band have vanished. The Fermi level is now situated above the completely filled valence band and LiNiO$_2$ is an
insulator, in agreement with experiment. Because SIC-LSD is still one-electron ground state theory it does not give
accurate removal energies of localized states, due to the unaccounted for screening and relaxation effects~\cite{walterPr}.
The localized states are therefore situated unphysically low at around -0.8 Ry (not shown).


Given the similarity of electronic structures of stoichiometric LiNiO$_2$ and NaNiO$_2$, what then is the origin of their very different properties? Although super-stoichiometric Li$_{1-x}$Ni$_{1+x}$O$_2$, $0 < x \leq 0.2$, was first synthesized by Dyer {\it et al.}~\cite{dyer} in 1953, attempts to synthesize the stoichiometric LiNiO$_2$ ($x=0$) compound have so far proved unsuccessful. There is now a growing body of evidence that chemical disorder inherent to 
super-stoichiometric LiNiO$_2$, but absent in NaNiO$_2$, is an important factor contributing to property 
differences. Indeed, in a recent theoretical paper, Mostovoy and Khomskii~\cite{mostovoy} raise the possibility that the complex electronic properties of LiNiO$_2$ are related to the presence of Ni anti-site ions in the Li layers. A number of experimental studies have also come to the same conclusion \cite{chappelPRB,barra}.
In super-stoichiometric Li$_{1-x}$Ni$_{1+x}$O$_2$ with excess Ni-atoms presumed to occupy the Li-layer, charge balance arguments suggest that misplaced Ni-atoms will have a $2+$ charge state ~\cite{nunez,barra} which will also induce an equivalent number of Ni$^{2+}$-ions in the NiO$_2$ layer. Namely, a valence state that can be captured by the crystallographic formula (Li$_{1-x}^{1+}$Ni$_x^{2+}$)$_{Li-layer}$(Ni$_x^{2+}$Ni$_{1-x}^{3+}$)$_{Ni-layer}$O$_2$. Despite its apparent complexity, the energetic basis of such a ground state can be investigated straightforwardly using the SIC-LSD method.
%

Starting from a supercell consisting of four LiNiO$_2$ formula units (crystal structure $R\bar{3}m$), we replace a single Li atom with Ni. In Table I, we compare the total energy for the trivalent scenario, i.e. (Li$_{3}^{1+}$Ni$^{3+}$)$_{Li-layer}$(Ni$_4^{3+}$)$_{Ni-layer}$O$_2$ in row 1, to a number of scenarios where an increasing number of Ni atoms is treated as divalent Ni$^{2+}$($d^8$). From Fig. \ref{lsddos}c we notice that in the energetically unfavorable, all Ni$^{3+}$ configuration, the substitution of a monovalent Li by a trivalent Ni results in additional e$_g$ bands that are partially filled. The gain in hybridization energy associated with this $d$-band formation is however not large enough to compete with the possible gain in self-interaction energy that results from localizing the corresponding $d$-states. Consequently, the energetically most favorable scenario is obtained with two Ni atoms in the Ni$^{2+}$ configuration and three Ni atoms in the Ni$^{3+}$ configuration (row 4 in Table I), i.e. (Li$_{3}^{1+}$Ni$^{2+}$)$_{Li-layer}$(Ni$_3^{3+}$Ni$^{2+}$)$_{Ni-layer}$O$_2$. The corresponding DOS is shown in Fig. \ref{lsddos}d, where it can be seen that, due to the localization of $d$-states, a large band gap is observed indicating that the compound remains insulating even when off-stoichiometric.

\begin{table}
\caption{Total energies for different Ni$^{3+}$/Ni$^{2+}$ configurations in Li$_3$Ni$_5$O$_8$, and Na$_3$MgNi$_4$O$_8$.
The energies are given relative to the corresponding all trivalent Ni$^{3+}$ configuration. The Ni$_{Li}$ notation refers to a Ni
substituted for a Li atom.}
\label{}
\begin{ruledtabular}
\begin{tabular}{|l|l|}
\multicolumn{2}{|c|}{Li$_3$Ni$_5$O$_8$}\\
\hline
Ni valency & Total energy (Ry) \\
\hline
Ni$_{Li}^{3+}$\;Ni$^{3+}$\;Ni$^{3+}$\;Ni$^{3+}$\;Ni$^{3+}$ & 0.00  \\
\hline
Ni$_{Li}^{3+}$\;Ni$^{3+}$\;Ni$^{3+}$\;Ni$^{3+}$\;Ni$^{2+}$ & -0.0393  \\
\hline
Ni$_{Li}^{2+}$\;Ni$^{3+}$\;Ni$^{3+}$\;Ni$^{3+}$\;Ni$^{3+}$ & -0.0691  \\
\hline
Ni$_{Li}^{2+}$\;Ni$^{3+}$\;Ni$^{3+}$\;Ni$^{3+}$\;Ni$^{2+}$ & -0.1033  \\
\hline
Ni$_{Li}^{2+}$\;Ni$^{3+}$\;Ni$^{3+}$\;Ni$^{2+}$\;Ni$^{2+}$ & 0.0262  \\
\hline
Ni$_{Li}^{2+}$\;Ni$^{2+}$\;Ni$^{2+}$\;Ni$^{2+}$\;Ni$^{2+}$ &0.1494   \\
\hline
\multicolumn{2}{|c|}{Na$_3$MgNi$_4$O$_8$}\\
\hline
Ni valency & Total energy (Ry) \\
\hline
Ni$^{3+}$\;Ni$^{3+}$\;Ni$^{3+}$\;Ni$^{3+}$ & 0.00  \\\hline
Ni$^{3+}$\;Ni$^{3+}$\;Ni$^{3+}$\;Ni$^{2+}$ & -0.0136 \\
\hline
Ni$^{3+}$\;Ni$^{3+}$\;Ni$^{2+}$\;Ni$^{2+}$ & +0.0658 \\\end{tabular}\end{ruledtabular}\label{NiforLi}
\end{table}

In the preceding discussion it is important to keep in mind that the valencies obtained in SIC-LSD are not ionic in nature. Rather, valency change is the result of a $d$-electron localization/delocalization transition and is not necessarily indicative of large charge transfer. Whether a given $d$-state will localize or take part in $d$-band formation depends on the outcome of the competition between hybridization and self-interaction energies.

The above interesting behavior of misplaced Ni-atoms notwithstanding, the question still remains as to the extent that the off-stoichiometry induced valence change can explain the observed differences in electronic and magnetic properties between LiNiO$_2$ and NaNiO$_2$.  Unlike Ni$^{3+}$ ions, divalent Ni$^{2+}$ ions are not JT-active. Consequently, a random distribution of these ions in the NiO$_2$ layer could possibly affect the orbital ordering. However, this issue is clouded by the fact that introduction of Ni ions into the Li layer gives rise to additional magnetic interactions between neighboring Ni layers which then need to be taken into account on top of the pre-existing intra-layer magnetic interactions. For example, it has been suggested that the substitution of some of the Na atoms by Ni atoms in NaNiO$_2$ leads to frustration of the antiferromagnetic stacking order and the destruction of long range magnetic order~\cite{lewis}. Given this complexity, it would clearly be advantageous to separate the effect of valency change on the orbital ordering and on the magnetic interactions. In what follows we suggest how this may be achieved.

Specifically, we propose to start from stoichiometric NaNiO$_2$, which is both magnetically and orbitally ordered, and substitute Mg or Ca
ions for some of the Na ions. By replacing monovalent Na with divalent Mg/Ca, we expect no additional interlayer
magnetic interactions to occur, but the extra valence electron will, dependent on the relative size of the corresponding self-interaction
and hybridization energies, either localize on a Ni-site or occupy an e$_g$ conduction band level.
Starting from a Na$_4$Ni$_4$O$_8$ unit cell and
replacing a single Na atom by Mg, we have studied a number of Ni-ion configurations. As shown in Table I, we find that the configuration with
three Ni ions in the 3+ charge state, and one Ni ion in the 2+ charge state is energetically most favorable. Thus, the substitution
of Mg/Ca atoms in the Na layer, leads to $d^{8}$ localization on an equivalent number of Ni-ions, and again results in a mixture of JT active
and non-active ions in the NiO$_2$ layer.  Clearly,
investigating experimentally how this would affect the local and medium range orbital ordering is then of considerable interest.

A study ~\cite{holz2} of mixed layered oxide phases Na$_{1-x}$Li$_{x}$NiO$_2$, where Li is gradually replaced by Na, has shown that both orbital-ordering and  antiferromagnetism are conserved for 0 $\leq$ x $\leq$ 0.2, and that the collective Jahn-Teller transition is only
suppressed for x $>$ 0.2. This result is in line with our picture, as the substitution of Na by isoelectronic Li does not result in a change in valency of the Ni ions in the NiO$_2$ layer. If the substitution of Mg or Ca for Na brought about the suppression of long range orbital ordering for comparatively smaller concentrations, this would be a strong indication that it is
indeed the random distribution of non JT-active Ni ions in the NiO$_2$ layer that is responsible, rather than the introduction of magnetic Ni atoms in the Li layers.
On the other hand if the orbital ordering persists up to higher Mg/Ca concentrations, it is more likely that the small ionic size of Li$^+$ is responsible for the suppression of macroscopic Jahn-Teller distortion.

In summary, we have applied the SIC-LSD method to study the electronic structure of both stoichiometric and off-stoichiometric nickelates.  Based on parameter free total energy considerations, we find the ground-states of both LiNiO$_2$ and NaNiO$_2$ to be insulating,
with Ni$^{3+}$ in the t$_{2g}^6$e$_g^1$ low-spin configuration. Furthermore, the calculations reveal that the crystal field effects rather than the exchange interactions determine the low spin ground state configuration.
With respect to Li$_{1-x}$Ni$_{1+x}$O$_2$, we find that for every Ni substituting for Li, a Ni-ion turns divalent in the NiO$_2$ layer. The presence of a divalent Ni-ion will alter the exchange interaction, and would strongly affect the magnetic properties of these compounds. We propose that NaNiO$_2$ having had Mg or Ca substituted for some of its Na would be ideally suited to experimental study of the effect of Ni$^{2+}$ ions on the long range orbital ordering.

\begin{acknowledgments}

This work was sponsored by the Laboratory Directed Research and Development Program (LDRD) program of
ORNL (LP, GMS, TE), and by the DOE-OS through the Offices of Basic Energy Sciences (BES),
Division of Materials Sciences and Engineering (LP, GMS, TE).
The calculations were carried out at the National Energy Research Scientific Computing Center (NERSC).
\end{acknowledgments}


\end{document}